\documentclass[aps,prd,twocolumn,nofootinbib,superscriptaddress,preprintnumbers]{revtex4}
\usepackage[hypertex]{hyperref}
\usepackage{graphicx}% Include figure files
\def\lesssim{\mathrel{\mathpalette\vereq<}}

\begin{document}

% Page numbers bottom-center
\pagestyle{plain}

\preprint{SLAC-PUB-11595}

\title{A Twin Higgs Model from Left-Right Symmetry}

\author{Z. Chacko}
\author{Hock-Seng Goh}
\affiliation{Department of Physics, University of Arizona, Tucson, AZ 85721}
\author{Roni Harnik}
\affiliation{
SLAC, Stanford University, Menlo Park, CA 94025}
\affiliation{Physics Department, Stanford University,
Stanford, CA 94305}

%\date{\today}

\begin{abstract}

We present twin Higgs models based on the extension of the Standard Model to
left-right symmetry that protect the weak scale against radiative corrections
up to scales of order 5 TeV. In the ultra-violet the Higgs sector of these
theories respects an approximate global symmetry, in addition to the discrete
parity symmetry characteristic of left-right symmetric models. The Standard
Model Higgs field emerges as the pseudo-Goldstone boson associated with the
breaking of the global symmetry. The parity symmetry tightly constrains the
form of radiative corrections to the Higgs potential, allowing natural
electroweak breaking. The minimal model predicts a rich spectrum of exotic
particles that will be accessible to upcoming experiments, and which are
necessary for the cancellation of one-loop quadratic divergences. These
include right-handed gauge bosons with masses not to exceed a few TeV and a
pair of vector-like quarks with masses of order several hundred~GeV.

\end{abstract}

 \maketitle

%%%%%%%%%%%%%%%%%%%%%%%%%%%%%%%%%%%%%%%%%%%%%%%%%%%%%%%%%%%%%%%%%%%%%%%%%%%%

\section{Introduction}

In the Standard Model (SM) the Higgs mass parameter receives quadratically
divergent quantum corrections that tend to destabilize the weak scale. This
suggests the existence of new physics near a TeV that resolves this problem.
However, precision electroweak measurements performed at LEP over the past
decade have lead to an apparent paradox~{\cite{LEPparadox}}. The problem is
that these experiments indicate
\begin{itemize}
\item
the existence of a light Higgs with mass less than about 250 GeV, and also that
\item
the cutoff $\Lambda$ for non-renormalizable operators that contribute to the precision
electroweak observables must be greater than about 5 TeV.
\end{itemize}
However Standard Model loop corrections from scales of order 5 TeV are
sufficiently large so as to generate a Higgs mass much larger than 250 GeV.
This is called the `LEP paradox'.  While we cannot rule out accidental
cancellations between different contributions to the LEP measurements, the
LEP paradox seems to suggest that whatever the new physics is that addresses
the hierarchy problem, it does not contribute significantly to the precision
electroweak observables. More concretely, there seem to be the three distinct
possibilities as follows.
\begin{itemize}
\item
There is no new physics below 5 TeV. In this case the Standard Model is simply
fine-tuned at the 2-3\% level or worse.
\item
The new physics which stabilizes the weak scale does contribute significantly
to precision electroweak observables, but satisfies the current bounds. In
this case the fact that the Standard Model with a light Higgs is a good fit
to the data is merely a coincidence.
\item
The new physics which stabilizes the weak scale does not contribute
significantly to precision electroweak observables.
\end{itemize}
Any complete solution to the LEP paradox should fall into the last category.
One such solution is weak scale supersymmetry, where R-parity suppresses
contributions to precision electroweak observables.

One interesting approach to the hierarchy problem, first proposed
in~\cite{GP, KG}, is that the Higgs mass parameter is protected against
radiative corrections because the Higgs is the pseudo-Goldstone boson of an
approximate global symmetry. In the last few years several interesting
realizations of this idea based on the little Higgs mechanism have been
constructed~{\cite{Little1, Little2}} (for a clear review and more references
see {\cite{Review}}).  These theories stabilize the weak scale up to 5 - 10
TeV. The underlying concept behind little Higgs theories is the idea of
`collective symmetry breaking' - the global symmetry is broken only when two
or more couplings in the Lagrangian are non-vanishing.  This is a significant
restriction on the form of the quantum corrections to the pseudo-Goldstone
potential, which can be used to realize natural electroweak symmetry
breaking. Models based on this idea where the corrections to precision
electroweak observables are small have been constructed {\cite{Tparity}},
(see also {\cite{custodial}}), and these naturally resolve the LEP paradox.

Recently twin Higgs models, an alternative class of realizations of the Higgs
as a pseudo-Goldstone boson, have been
proposed~{\cite{twin,BGH,CNPP}}.  These theories also
protect the weak scale from radiative corrections up to scales of order 5 -
10 TeV, but in a manner completely distinct from little Higgs theories.  In
the ultra-violet these theories respect a discrete $Z_2$ interchange symmetry
in addition to an approximate global symmetry of the Higgs sector.  The
Standard Model Higgs field emerges as the pseudo-Goldstone boson associated
with the breaking of the global symmetry. The discrete symmetry is enough to
ensure that any quadratically divergent contribution to the Higgs potential
accidentally respects the global symmetry. The pseudo-Goldstone mass is then
at most logarithmically divergent, allowing natural electroweak breaking to
be realized. Corrections to precision electroweak observables can be
naturally small, providing a resolution of the LEP paradox. In the original
incarnation of this idea the discrete symmetry corresponded to the
interchange of every SM particle with the corresponding particle transforming
under a mirror SM~\cite{twin}. These models have the intriguing feature that
all of the new physics beyond the SM (or beyond a minimal extension of the
SM~\cite{CNPP}) is a singlet under the SM gauge groups. Such new physics will
then appear in upcoming experiments purely as missing energy, posing an
interesting challenge for the LHC.

In this paper we present a more minimal realization of the twin Higgs
mechanism that does not involve adding a whole new mirror copy of the SM.
Instead, we identify the discrete symmetry with the parity symmetry
associated with the extension of the SM to a left-right symmetric
model~{\cite{LR}}. This directly leads to a class of interesting models with
exciting implications for upcoming experiments.

This paper is organized as follows. In Section~\ref{toy} we illustrate these
ideas by presenting a simple model where the symmetries are realized
linearly. In Section~\ref{nonlinear} we present a more general non-linear
realization and demonstrate that natural electroweak breaking can be
obtained. In Section~\ref{pheno} we briefly discuss some phenomenological
aspects of our model and summarize.

\section{A Linear Realization}

\label{toy}
We illustrate how the symmetries are implemented in these models by considering
first a simplified model where the global symmetry is realized linearly.
Consider a complex scalar field, $H$, that transforms as a fundamental under a
global $U(4)$ symmetry. The potential for this field is given by
\begin{equation}
\label{treepotential}
    V(H)= -m^2 H^\dagger H + {\lambda}{}(H^\dagger H)^2\,.
\end{equation}
Since the mass squared of $H$ is negative it will develop a VEV, $\langle |H|
\rangle = m/\sqrt{2\lambda}\equiv f$, that breaks $U(4)\to U(3)$ yielding 7
massless Nambu-Goldstone bosons.  We now break the $U(4)$ explicitly by
gauging an $SU(2)_L\times SU(2)_R$ subgroup. Here $SU(2)_L$ generates the
weak interactions of the SM, while $SU(2)_R$ generates the corresponding
right-handed interactions associated with the extension of the SM to the
left-right symmetric model. (We defer a discussion of the $U(1)_{B-L}$ gauge
symmetry.)  The field $H$ transforms as
\begin{equation}
	H=\pmatrix{H_L \cr H_R}
%H=\left({ H_L , H_R} \right)
\end{equation}
where
$H_L$ is a doublet under $SU(2)_L$ that is to be identified with the SM Higgs
and $H_R$ is a doublet under $SU(2)_R$.  This Higgs structure is
characteristic of Alternative Left-Right Symmetric Models {\cite{ALRM}}, (see
also {\cite{RabiLR}}).

Since $U(4)$ is now broken explicitly, we expect that the would-be Goldstones
pick up a mass that is proportional to the explicit breaking.  Specifically,
gauge loops contribute a quadratically divergent mass to the components of
$H$ as
\begin{equation}
    \Delta V= \frac{9 g_L^2 \Lambda^2}{64\pi^2}H_L^\dagger H_L
        + \frac{9 g_R^2 \Lambda^2}{64\pi^2}H_R^\dagger H_R +\ldots\,,
\end{equation}
a loop factor below the cutoff $\Lambda$ of the theory.  If we now impose
parity symmetry the two gauge couplings have to be equal,
$g_L=g_R\equiv g$, so that
\begin{equation}
    \Delta V= \frac{9 g^2 \Lambda^2}{64\pi^2}
    (H_L^\dagger H_L + H_R^\dagger H_R)=
    \frac{9 g^2 \Lambda^2}{64\pi^2} H^\dagger H
\end{equation}
which is invariant under $U(4)$ and therefore will not contribute a mass to
the Goldstones.  In other words, left-right symmetry constrains the
quadratically divergent mass terms to have a $U(4)$ invariant form.  The
Goldstones are therefore completely insensitive to quadratic divergences from
gauge loops.

Gauge loops will however contribute a logarithmically divergent term to the
potential that is not $U(4)$ symmetric and has the general form $ \kappa
\left(|H_L|^4 + |H_R|^4 \right)$ where $\kappa$ is of order $g^4/16 \pi^2
{\rm log} \left(\Lambda/ gf \right)$. Provided $\Lambda$ is not very much
larger than $f$ this leads to the would-be Goldstones acquiring a mass of
order $g^2 f/4 \pi$ which is of order the weak scale for $f$ of order a TeV.

At this point we note that the Higgs potential of Eq.~({\ref{treepotential}})
actually possesses a larger global $O(8)$ symmetry of which $U(4)$ is merely
a sub-group, and the 7 Goldstone bosons we have identified can also be
thought of as emerging from the breaking of $O(8)$ to $O(7)$. In particular,
this $O(8)$ symmetry includes the custodial $SU(2)$ of the Higgs potential in
the Standard Model.

This approach to stabilizing the weak scale against quantum corrections from
gauge loops can be generalized to include all the other interactions in the SM
by making the entire theory left-right symmetric. The fermionic content of the
theory is then three generations of
\begin{eqnarray}
Q_L = \left(u,d \right)_L =\left[2,1,1/3 \right] \; \;
L_L = \left(\nu,e \right)_L = \left[2,1, -1 \right] \nonumber \\
Q_R = \left(u,d \right)_R = \left[1,2, 1/3 \right] \; \;
L_R = \left(\nu,e \right)_R = \left[1,2, -1 \right]
\end{eqnarray}
where the square brackets indicate the quantum numbers of the corresponding
field under $SU(2)_L \times SU(2)_R \times U(1)_{B - L}$. We see that in
addition to the SM fermions the theory includes right-handed neutrinos as
required by left-right symmetry. The Higgs fields have quantum numbers
\begin{equation}
H_L = \left[2, 1, 1
\right] \; \; \; \; \; \; H_R = \left[1,2,1 \right]
\end{equation}
under $SU(2)_L \times SU(2)_R \times U(1)_{B - L}$.
The down-type Yukawa couplings of the SM emerge from non-renormalizable
couplings of the form
\begin{equation}
\left(\frac{\overline{Q}_R H_R H_L^{\dagger} Q_L \; \; + \;\;
\overline{L}_R H_R H_L^{\dagger} L_L}{\Lambda} \right)  \; \; + \; \;
{\rm h.c.}
\end{equation}
The up-type Yukawa couplings of the SM emerge from non-renormalizable couplings
of the form
\begin{equation}
\left(\frac{\overline{Q_R} \; H_R^{\dagger} H_L Q_L \; \; + \;\;
{\rm h.c.}}{\Lambda} \right)
\end{equation}
When the field $H_R$ acquires a VEV of order $f$ breaking $SU(2)_R \times
U(1)_{B - L}$ down to $U(1)_Y$ these non-renormalizable couplings reduce to
the familiar Yukawa couplings of the SM. Unfortunately, although this works
well for the smaller Yukawa couplings, it is not satisfactory for the top
Yukawa coupling which is required to be order one. We address this difficulty
by introducing a vector-like pair of quarks $T_L$ and $T_R$ which have the
quantum numbers
\begin{equation}
T_L = \left[1, 1, 4/3
\right] \; \; \; \; \; \; T_R = \left[1,1, 4/3 \right]
\end{equation}
under $SU(2)_L \times SU(2)_R \times U(1)_{B - L}$. We can then write the
interactions
\begin{equation}
\label{topmodule}
\left( y \; \overline{Q_R} \; H_R^{\dagger} T_L \; + \;
y \; \overline{Q_L} \; H_L^{\dagger} T_R \; + \;
M \overline{T_L} T_R \right)  \; \; + \; \; {\rm h.c.}
\end{equation}
The right-handed top quark of the SM then emerges as a linear combination of
$T_R$ and the third generation up-type quark in $Q_R$, while the orthogonal
linear combination is heavy. Provided $M \lesssim f$ and $y$ is of order one
the physical top Yukawa will then also be of order one. The parameter $M$
controls the mixing of the left-handed top with the SU(2$)_L$ singlet $T_L$ ,
and is therefore constrained by $Z \rightarrow b \; \bar{b}$. However,
nothing prevents $M$ from simply being set to zero and therefore this is not
a particularly tight constraint.

The fact that the entire theory is now left-right symmetric ensures that any
quadratically divergent contribution to the Higgs mass has a form $\propto
\Lambda^2 (|H_L|^2 + |H_R|^2)$ which is harmless due to its accidental $U(4)$
symmetry. Although quantum corrections to the quartic are in general not
$U(4)$ invariant, once again these only lead to logarithmically divergent
contributions to the mass of the pseudo-Goldstone Higgs field, allowing for a
natural hierarchy between $f$ and the weak scale.

Unfortunately the theory as described above is still not entirely
satisfactory.  The reason is that precision electroweak constraints on
$SU(2)_R$ gauge bosons force the scale $f$ to lie close to 2 TeV or
above~{\cite{Kingman}, which tends to reintroduce fine-tuning. While there
may be several possible solutions to this problem, for the remainder of this
paper we shall concentrate on only one.  
%%%%%%
%%%%%%
A key observation is that if $f$ indeed lied at 2~TeV, the fine
tuning from SM gauge loops would still be milder compared to the fine tuning
that arises from the top with $f\sim500-800$~GeV. We would thus like to
raise the effective symmetry breaking scale in the gauge, but not the top
sector. We can indeed do just that by
%%%%%%%%%%%
%%%%%%%%%%%
introducing an additional Higgs field $\hat{H} = ( \hat{H}_L , \hat{H}_R )$
into the theory, where $\hat{H}_L$ and $\hat{H}_R$ have exactly the same
gauge quantum numbers as $H_L$ and $H_R$, but do not have the corresponding
couplings to the SM fermions.  We assume that $\hat{H}$ and $H$ do not couple
directly to each other at the scale $\Lambda$, and further that the potential
for $\hat{H}$ at this scale has the $U(4)$ invariant form 
\begin{equation}
V(\hat{H})= -\hat{m}^2 \hat{H}^\dagger \hat{H}+\hat{\lambda}{}(\hat{H}^\dagger
\hat{H})^2\,.  \end{equation} Then the Higgs sector of the theory has an
approximate $U(4) \times U(4)$ symmetry, or more precisely an approximate
$O(8) \times O(8)$ symmetry, of which the $SU(2)_L \times SU(2)_R \times
U(1)_{B - L}$ sub-group is gauged.  If $\hat{H}_R$ acquires a VEV $\hat{f} >
2$ TeV breaking $SU(2)_R \times U(1)_{B - L}$ to $U(1)_Y$ then the precision
electroweak constraints on this theory from the extra gauge bosons are
satisfied. At the same time, the approximate $U(4) \times U(4)$ symmetry
implies that the twin symmetric form of the potential for $H$ is not
significantly affected, so that electroweak symmetry breaking can still occur
naturally.

\section{A Non-Linear Realization}\label{nonlinear}

We now construct a realistic twin symmetric model that implements these
symmetries non-linearly. The pseudo-Goldstone fields of the non-linear model
are those which survive after integrating out the radial modes of the fields
$H$ and $\hat{H}$ in the linear model. We parameterize these degrees of
freedom as
\begin{eqnarray}
\label{pseudo}
H &=& \exp(\frac{i}{f} {h}^a t^a) \pmatrix{ 0 \cr 0 \cr 0 \cr f }
\equiv \pmatrix{ 0 \cr 0 \cr 0 \cr f }
+ i \pmatrix{ h^1 \cr h^2 \cr h^3 \cr h^0 } + \ldots
\nonumber \\
&&
\\
\hat{H} &=& \exp(\frac{i}{\hat{f}} {\hat{h}}^a t^a) \pmatrix{ 0 \cr 0 \cr 0 \cr
\hat{f}
}
\equiv \pmatrix{ 0 \cr 0 \cr 0 \cr \hat{f} }
+ i \pmatrix{ \hat{h}^1 \cr \hat{h}^2 \cr \hat{h}^3 \cr \hat{h}^0 } + \ldots
\nonumber
\end{eqnarray}
where $h^{1,\ldots,3}$, $\hat{h}^{1,\ldots,3}$ are complex and $h^0$,
$\hat{h}^0$ are real. The $t^a$ are a suitably chosen set of broken
generators.  In general the effective theory for these fields will contain
all of the operators allowed by the non-linearly realized $U(4) \times U(4)$
symmetry, suppressed by the cutoff scale $\Lambda$. However, in order to
suppress custodial SU(2) violation we assume that the symmetry which is
non-linearly realized is in fact $O(8) \times O(8)$. This provides additional
restrictions on the form of the interactions in the effective theory below
$\Lambda$, allowing precision electroweak constraints from higher dimensional
operators to be naturally satisfied.  If the theory is strongly coupled at
the cutoff we can estimate $\Lambda\sim 4 \pi f$. However, we do not exclude
the possibility that $\Lambda$ is less than this.  For example, if the UV
completion of the non-linear model is the linear model then $\Lambda$ is
simply the mass of the radial mode.

In general, any potential for the pseudo-Goldstone fields can only emerge from those
interactions which violate the global symmetries, specifically the gauge and Yukawa
couplings. In particular the electroweak gauge interactions and the top Yukawa
contribute the most to the pseudo-Goldstone Higgs potential and must therefore be
studied in detail. We will therefore calculate the contributions to the one loop
Coleman-Weinberg (CW) potential~\cite{CW} from these couplings. At one loop the gauge
and top sectors contribute separately, simplifying the calculation.

As before, we gauge the $SU(2)_L \times SU(2)_R \times U(1)_{B - L}$
sub-groups of the global symmetry. The VEVS $f$ and $\hat{f}$ break
$SU(2)_R\times U(1)_{B - L}$ down to $U(1)_Y$, giving $W_R$ and $Z_R$ masses
of order $g \hat{f}$. One linear combination of the fields $H_R$ and
$\hat{H}_R$ is eaten.  The $SU(2)_L$ doublet $h^T\equiv (h^1, h^2)$ is left
uneaten and is identified as the SM Higgs. The couplings of the
pseudo-Goldstone fields to the gauge fields is given by expanding out $H$ and
$\hat{H}$ in terms of the pseudo-Goldstones as given by eq. ({\ref{pseudo}})
in the interaction
\begin{eqnarray}
&&\left[ \left|\left(\partial_\mu + ig W_{\mu, L}
+ \frac{i}{2} g' B_{\mu}\right) H_L \right|^2
+ \left( L \rightarrow R \right) \right] + \nonumber \\
&&\left[ \left|\left(\partial_\mu + ig W_{\mu, L}
+ \frac{i}{2} g' B_{\mu}\right) \hat{H}_L \right|^2
+ \left( L \rightarrow R \right) \right]
\end{eqnarray}
where $B_{\mu}$ is the gauge boson of $U(1)_{B - L}$.  A simple way of
calculating the effective potential is to determine the vacuum energy as a
function of the field dependent masses of all of the fields in the theory.
In the absence of quadratic divergences this leads to the formula
\begin{equation}
    V_{CW} = \pm \frac{1}{64\pi^2} \sum_i {M_i^4}
    \left(\log\frac{\Lambda^2}{M_i^2} + \frac{3}{2}\right)
\end{equation}
where the sum is over all degrees of freedom, the sign being negative for
bosons and positive for fermions.  Writing the Higgs potential in the form
\begin{equation}
\label{potential}
        V(h)= m_h^2 h^\dagger h + \lambda_h (h^\dagger h)^2 + \ldots
\end{equation}
we find that the contribution to the Higgs mass term from the gauge sector is
\begin{eqnarray}
\label{gauge}
        m_h^2|_{gauge}
    &=& \frac{3 g^2 M^2_{W_R}}{32\pi^2 }
    \left( \log\frac{ \Lambda^2} {M^2_{W_R}} + 1 \right) \\
    &+& \frac{3 g^2 (2M^2_{Z_R}-M^2_{W_R})}{64\pi^2 }
    \left( \log\frac{ \Lambda^2} {M^2_{Z_R}} + 1 \right) \,
    \nonumber\\
    \lambda_h|_{gauge} &=& -\frac{m_h^2|_{gauge}}{3f^2}\nonumber
\end{eqnarray}
where $M_{W_R}^2= g^2 (f^2+\hat{f}^2)/2$ and $M_{Z_R}^2=
(g^2+g'^2)(f^2+\hat{f}^2)/2 \,$.
Except the term proportional to the Higgs mass squared, all other
contributions to the Higgs quartic from this sector are small and can be
neglected.

We now turn to the top sector. The couplings of the pseudo-Goldstone fields to
the top quark are obtained by expanding out $H$ as in eq. ({\ref{pseudo}}) in
the interactions of eq. ({\ref{topmodule}}) which generate the top Yukawa
coupling.  The $h$ dependent masses of the fields in the top sector are
determined from this and can be expressed as
\begin{eqnarray}
       m^2_{Q}=\frac{y^4 f^2}{M^2 +y^2f^2} h^\dagger h &\qquad&
       m^2_{T}= M^2 +y^2f^2 \nonumber\\
%       m^2_{t_B}= y^2 f^2 \qquad\qquad &\qquad&
%       m^2_{T_B}= M^2
\end{eqnarray}
to leading order in $|h|^2/f^2$, where we have assumed for simplicity that
$y$ is real. This leads to the following contributions to the Higgs potential
of eq.~({\ref{potential}}).
\begin{eqnarray}
\label{topcont}
        m_h^2|_{top}&=&
        -\frac{3}{8\pi^2}y^2_t m_{T}^2\left(
        \log\frac{\Lambda^2}{m^2_{T}}+1\right)\,,
        \nonumber\\
        \lambda_h|_{top} &=& -\frac{m_h^2|_{top}}{3f^2}+
        \frac{3}{16 \pi^2}\left(y_t^4
        \log\frac{m^2_{T}}{m^2_{Q}}+ 2 y^4 \log\frac{\Lambda^2}{m^2_{T}}\right)\nonumber\\
        &&- \frac{3 }{32\pi^2}
        \left[y_t^4
        -4y^4\right]
\end{eqnarray}
where $y_t$ is defined by
\begin{equation}
y_t = \frac{y^2 f}{\sqrt{M^2 +y^2f^2}}
\end{equation}
This completes the determination of the one-loop potential for the SM Higgs.
One may worry that corrections to the SM Higgs mass squared of order $g^2
f^2$ may arise at higher loop order~{\cite{BGH}}. However, we show in the
appendix that this is not the case.

It is also necessary to show that the other pseudo-Goldstone fields in $H$
and $\hat{H}$ also have positive mass squareds. It is straightforward to
ensure that $\hat{H}_L$ has positive mass squared by adding to the potential
a term $\hat{\mu}^2 {\hat{H}^\dagger}_L \hat{H}_L$ where $\hat{\mu}$ is of
order ${f}$. Such a term breaks both parity and the approximate $U(4)$
symmetry of the potential for $\hat{H}$, but only softly. It is therefore
technically natural for $\hat{\mu}$ to be smaller than $\Lambda$.

What about the fields in $H_R$ and $\hat{H}_R$? Of these six fields, three
are eaten and become the longitudinal components of the right-handed gauge
bosons while the remaining three remain light as pseudo-Goldstone bosons
associated with the breaking of the approximate $U(2)_R \times U(2)_R$
symmetry of the Higgs potential. Of the light fields, two carry electric
charges of +1 and -1 while the last is neutral. The electrically charged
fields acquire positive mass squareds from the gauge interactions which
violate $U(2)_R \times U(2)_R$, but the neutral state remains massless. In
order to give it a mass we add to the potential a term $B \; H_R^{\dagger}
\hat{H}_R$ where $\sqrt{B} $ is of order 50 -- 100 GeV or so.  Since this is
the only term in the Lagrangian which breaks the discrete symmetry $\hat{H}_R
\rightarrow - \hat{H}_R$ it is technically natural for it to be small.

In this non-linear model, the absence of quadratically divergent contributions
to the Higgs mass can be understood as a consequence of cancellations between
the familiar SM loop corrections and new loop corrections that arise from the
(mostly non-renormalizable) couplings of the Higgs to the twin sector.

We are now in a position to estimate the fine-tuning in this class of models.
Unfortunately, for a fixed value of the cutoff $\Lambda$, the precision of
this estimate is necessarily limited by the fact that the answer is sensitive
to the exact relation between $f$ and $\Lambda$, which in a strongly coupled
theory depends both on the detailed dynamics of the theory and also on the
physical observable under consideration.  A naive estimate gives $\Lambda
\sim 4 \pi f$. However, it was shown in {\cite{BGH}} that in the linear
model, in the absence of the radial mode, unitarity is saturated at $\Lambda
\sim 2 \pi f$. Therefore, in order to get some sense of the fine-tuning we
will allow for both possibilities, considering points in parameter space
satisfying the relation $\Lambda = 4 \pi f$ as well as points in parameter
space satisfying the relation $\Lambda = 2 \pi f$.

For $f = 800$ GeV, $\Lambda \sim 4 \pi f \approx 10$ TeV, $M =
150$ GeV, $\sqrt{B} = 50$ GeV we find that in order to obtain the
SM values of $M_W$ and $M_Z$ we need $\hat{f} \approx$ 4.29 TeV.
The Higgs mass is then about 174 GeV. Estimating the fine-tuning
as $\partial \; {\rm log} M_Z^2 / \partial \; {\rm log} f^2 $ we
find that it is of order 12\% (1 in 8). Similarly for $f = 800$
GeV, $\Lambda \sim 2 \pi f \approx 5$ TeV, $M = 150$ GeV,
$\sqrt{B} = 50$ GeV we find that in order to obtain the SM values
of $M_W$ and $M_Z$ we need $\hat{f} \approx$ 4.68 TeV. The Higgs
mass is then about 155 GeV. Estimating the fine-tuning as
$\partial \; {\rm log} M_Z^2 /
\partial \; {\rm log} {f}^2 $ we find that it is of order 12\% (1 in 8).
These and other results are summarized in Table~\ref{table-summary}. This
shows that these models stabilize the weak scale up to about 5 TeV.

%%%%%%%%%%%%%%%%%%%%%%%%%%%%%
\begin{table}[tbh]
\label{table1}
\begin{tabular}{|c|c|c|c|c|c|c|} \hline $\Lambda${\tiny (TeV)} & $f${\tiny (GeV)} &
$\hat{f}${\tiny (TeV)} & $M${\tiny (GeV)} & $\sqrt{B}${\tiny(GeV)}
& $m_{h}${\tiny (GeV)} & Tuning \\ \hline 10 & 800 & 4.29 & 150 &
50 & 174 & 0.117 \\ 6 & 500 & 2.27 & 150 & 50 & 172 & 0.270
\\ 5 & 800 & 4.68 & 150 & 50 & 155 & 0.124
%\\ \hline 10 & 800 & 4.25 & 150 & 50
%& 177 & 0.119 \\ 6 & 500 & 2.23 & 150 & 50 & 180 & 0.268
%\\ 10 & 800 &
%--- & 0 & 355 & 166 & 0.112 \\ 6 & 500 & --- & 0 & 203 & 153 &
%0.307
\\ \hline \end{tabular} \caption{ A summary of the Higgs mass
and fine tuning, $\partial \; {\rm log} M_Z^2 / \partial \; {\rm
log} {f}^2 $, for sample points of parameter
space. The largest fine tuning is associated with ${f}$.}
\label{table-summary}
\end{table}
%%%%%%%%%%%%%%%%%%%%%%%%%%%%%%%%%%%%%%%%%%%%%

To what extent does the absence of a tree-level quartic affect the
fine-tuning in these theories? To understand this, consider a theory with a
single light Higgs doublet at low energies and a scalar potential of the form
\begin{equation}
\label{potential}
        V(h)= m_h^2 h^\dagger h + \lambda_h (h^\dagger h)^2
\end{equation}
In terms of these parameters the electroweak VEV $v = \sqrt{|m_h|^2/2
\lambda_h}$ and the physical Higgs mass, which we denote by $m_{h, {\rm
phys}}$, is given by $\sqrt{2}|m_h|$. In our model, the dominant contribution
to the Higgs mass parameter $m_h^2$ arises from the top Yukawa coupling.  If
we denote this contribution by $m_h^2|_{{\rm top}}$, a good estimate of the
fine-tuning may be obtained by considering the ratio $m_h^2/m_h^2|_{{\rm
top}}$. This is equal to $m_{h, {\rm phys}}^2/(2 \; m_h^2|_{{\rm top}})$.
Now, is clear from Table~[I] that $m_{h, {\rm phys}}$ in our models is of
order 150 GeV or larger. Since precision electroweak constraints require the
Higgs to be lighter than about 250 GeV, the potential improvement in
fine-tuning in our model from a tree-level quartic in the region of parameter
space where the LEP paradox is addressed is at most of order $(250)^2/(150)^2
\approx 3$, and close to 2 for most of the points in the table.

The preceding analysis enables us to compare the fine-tuning in our model to
that in little Higgs theories with a tree level quartic. For concreteness we
focus on the little Higgs model of Kaplan and Schmaltz~{\cite{double}}, for
which the pattern of symmetry breaking, SU(4$)^4 \rightarrow $ SU(3$)^4$, is
most similar to ours, and for which $f$ and $\Lambda$ can therefore be
defined in close analogy.  In this theory the low energy spectrum contains
two Higgs doublets, of which only one couples to the top quark. In the limit
where this doublet is significantly lighter than the other we can obtain a
simple estimate of the fine-tuning. To do this we calculate $m_h^2|_{{\rm
top}}$ for the light doublet, and compute $m_{h, {\rm phys}}^2/2 \;
(m_h^2|_{{\rm top}})$, setting $m_{h, {\rm phys}}$ to its upper bound of 250
GeV. The gauge symmetry is SU(3$)_C$ $ \times$ SU(4$)$ $\times $ U(1$)_X$,
where SU(4$) \times $U(1$)_X$ is broken down to the familiar SU(2$)_L$ $
\times $ U(1$)_Y$ of the Standard Model.  The top Yukawa coupling emerges
from couplings of the form
\begin{equation}
\left[ y_1 \overline{Q} H t_1 + y_2 \overline{Q} \hat{H} t_2 + {\rm h.c.}
\right]
\end{equation}
The gauge quantum numbers of these fields under SU(3$)_C$ $ \times$ SU(4) are $Q \equiv
[3,
{4}]$, $t_1, t_2 \equiv [3, 1]$ and $H, \hat{H} \equiv [1, 4]$. The third generation
quark
doublet of the Standard Model is contained in $Q$ while the right-handed top quark
emerges
from a linear combination of $t_1$ and $t_2$. The light Higgs doublet emerges as the
uneaten linear combination of
the doublets in $H$ and $\hat{H}$, which may be expanded out exactly as in
Eq.~({\ref{pseudo}}).
Then a simple calculation {\cite{Simplest}} shows that for this theory, the divergent 
part of
$m_h^2|_{top}$ is bounded from below as
\begin{equation}
{\label{KSformula}}
|m_h^2|_{top}| \ge 2 \; \frac{3 y_t^2 f^2}{8 \pi^2}{\rm log}\frac{\Lambda^2}{f^2}
\end{equation}
where we have assumed $f < \hat{f}$ without loss of generality. This must be
compared against the contributions to $m_h^2|_{{\rm top}}$ in the twin Higgs
model as given by Eq.~(\ref{topcont}). We see that for small $M$, assuming
fixed values of $f$ and $\Lambda$, the value of $m_h^2|_{{\rm top}}$ in the
little Higgs model is larger by a factor of 2 or more.  From this it follows
that in spite of the absence of a tree-level quartic in our model, for fixed
values of $f$ and $\Lambda$, the fine-tuning in the two models is in fact
quite comparable. However precision electroweak constraints on the twin Higgs
model are much weaker due to the absence of SU(4) gauge bosons, which means
that much lower values of $f$ are experimentally allowed than in this little
Higgs model. This translates to a significant improvement in fine-tuning over
the little Higgs case. Some models based on collective symmetry breaking
where the bounds from precision electroweak measurements are weaker and which
admit low values of $f$ have been constructed, for example, {\cite{Tparity}},
{\cite{lowf}}. A study of the relative fine-tuning with respect to these
models is left for future work.

\section{Phenomenology}\label{pheno}

These models predict a rich spectrum of light exotics which can be detected in
the next generation of collider experiments. These include
\begin{itemize}
\item
the right-handed gauge bosons, which have masses
not to exceed a few TeV, and which couple with the same strength as the gauge
bosons of $SU(2)_L$,
\item
the vector-like quarks $T_L$ and $T_R$ which are expected to have masses
of several hundred GeV
\item
the charged pseudo-Goldstones from $H_R$ and $\hat{H}_R$,
which have masses not to exceed a few
hundred GeV
\item
the neutral pseudo-Goldstone from  $H_R$ and $\hat{H}_R$ which
also has mass of order a hundred GeV
\end{itemize}
A detailed study of the collider signatures of this model is left for future
work.  Since $\hat{H}_L \rightarrow - \hat{H}_L$ is an exact symmetry of the
model the neutral component of this field is a natural dark matter candidate.

We now turn to the question of how neutrino masses are generated in this
model. Dirac neutrino masses arise from the operator
$[ (\overline{L}_R H_R^{\dagger} H_L L_L/ \Lambda)
\; + \;
{\rm h.c.} ]$
while the operator
$[ ({{L}_R  \hat{H}_R \hat{H}_R L_R)/\Lambda \; + \;
{\rm h.c.}}]$
generates a Majorana mass for the right-handed neutrinos. This allows the SM
neutrinos to get a Majorana mass of the right size through the see-saw mechanism
{\cite{seesaw}}, provided the coefficient of the operator which generates the
Dirac neutrino mass is small $\sim 10^{-5}$.

%\section{Conclusions}

In summary we have constructed a new class of twin Higgs models based on
parity-symmetric left-right models which stabilize the weak scale
against radiative corrections up to scales of order 5 TeV. These theories
make definite predictions for exotic particles that can be detected in the
next generation of collider experiments, and admit a natural dark matter
candidate.
\\

%\acknowledgments

\noindent {\bf Acknowledgments --}
We thank Spencer Chang and Yasunori Nomura for discussions. Z.C and H.S.G are
supported by the NSF under grant PHY-0408954.  R.H is supported by the US
Department of Energy under Contract No.~DE-AC02-76SF00515.

\appendix
\section{2 Loop Diagrams}\label{appendixA}

In~\cite{BGH} it was pointed out that in the strong coupling limit of the
linear model one expects two loop contributions to the $U(4)$ violating
quartic, $|H_L|^4 + |H_R|^4$, of order $g^2$. One can see that such
contributions might be present by inspecting 2-loop diagrams such as those of
Figure~\ref{graphs} that are of order $g^2 \lambda^2/(16\pi^2)^2$ and taking
$\lambda$ to its NDA value of $(4\pi)^2$. The presence of such a large
quartic would imply that in the strong coupling limit the Higgs mass in our
model (and in the model of~\cite{twin}) is close to the upper bound allowed
by precision electroweak data. Moreover the exact Higgs mass would be
uncalculable. In this appendix we show that summing all of the relevant
2-loop graphs in fact yields a $U(4)$ symmetric quartic. In the next appendix
we show that this is not an accident but can be understood as a consequence
of a symmetry argument which can be extended to all loop orders.

The 2-loop diagrams that will potentially contribute at order $g^2$ may be
divided into four subcategories as follows: (a)~the gauge boson connects
between two external legs, (b)~the gauge boson connects an internal and an
external scalar leg, (c)~the gauge boson connects to a common internal leg,
thus correcting the scalar propagator, and (d)~the gauge boson connects
between two different internal legs. A representative of each group is shown
in Figure~\ref{graphs}.  Because only the $SU(2)_L\times SU(2)_R$ subgroup of
$U(4)$ is gauged, we only consider diagrams where the gauge bosons are
exchanged between a pair of fields labeled $L$ or $R$ and not when the
exchange is between $L$ and $R$.  Note that we can ignore contribution form
$B-L$ gauge boson exchange because the $B-L$ charges of the Higgs fields
respect $U(4)$.

\begin{figure}[t!]
\begin{tabular}{ccc}
\includegraphics[width=0.45\columnwidth]{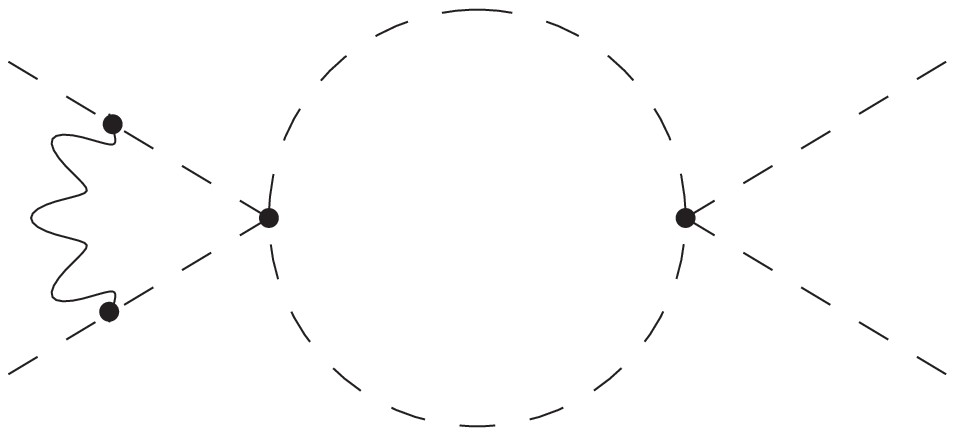} & \ &
\includegraphics[width=0.45\columnwidth]{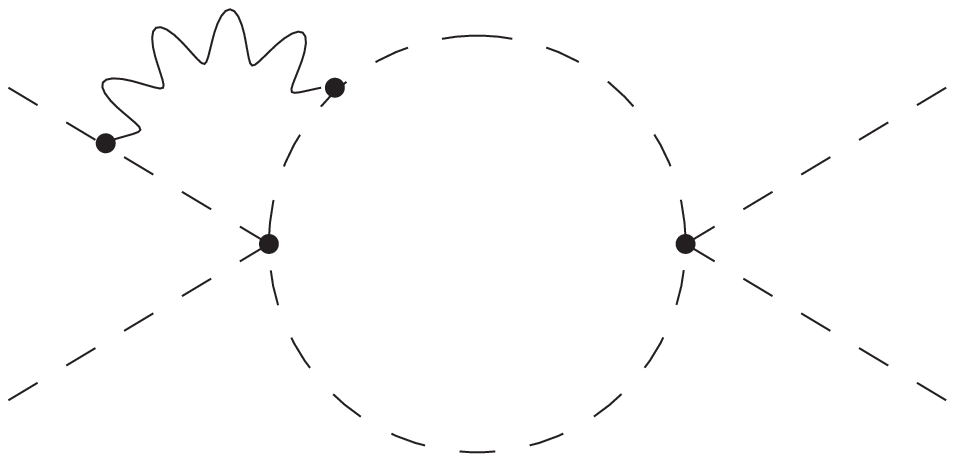}\\
{\small (a)} & \ & {\small (b)}\\ \ & \ & \ \\
\includegraphics[width=0.45\columnwidth]{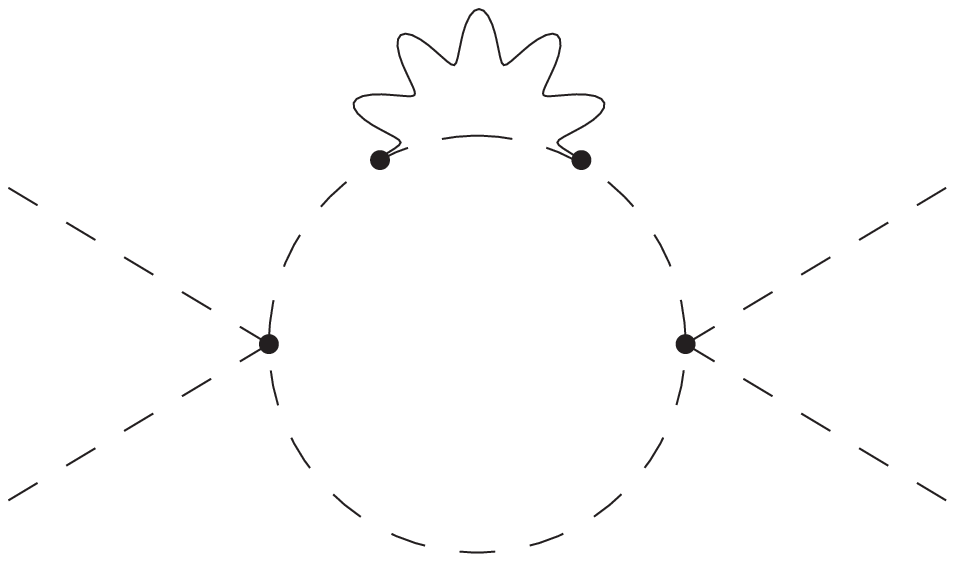} & \ &
\includegraphics[width=0.45\columnwidth]{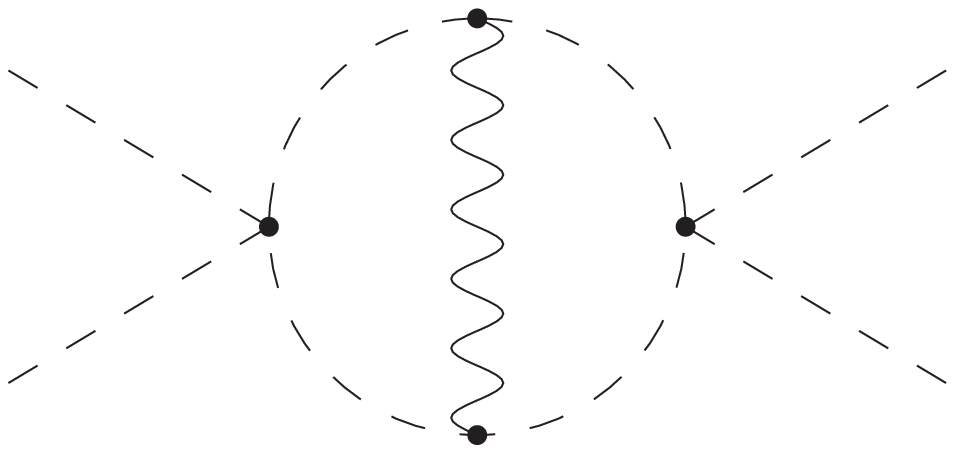}\\
% \ & \ & \ \\
{\small (c)} & \ & {\small (d)}
\end{tabular}
\caption{Representative graphs that contribute at order
$g^2\lambda^2/(16\pi^2)^2$. Summing all such diagrams yields an $SU(4)$
symmetric quartic.}\label{graphs}
\end{figure}

To show that the 2-loop quartic is $U(4)$ symmetric we can factor a common
phase space integral from each category of diagrams and focus our attention
to symmetry factors, signs, and group theory coefficients. In particular, a
relative sign difference between diagrams arises when a gauge boson is
exchanged between two $H$'s versus the case where it is exchanged between an
$H$ and an $H^\dagger$. In order to show that the quartic is $U(4)$ symmetric
we need to show that the $|H_L|^2|H_R|^2$ quartic is twice as large as the
$|H_L|^4$.

The quartics arising from graphs of type (a) are 1-loop corrections to to an
already $U(4)$ symmetric (1-loop) quartic due to the derivative interaction
of the gauge boson. It is then sufficient to show that the $|H_L|^4$ and
$|H_L|^2|H_R|^2$ are renormalized in the same way. There are 6 diagrams with
identical kinematics that contribute to $|H_L|^4$, however two of them have a
relative minus sign giving a total symmetry factor of $4-2=2$. The
$|H_L|^2|H_R|^2$ receives contribution from just two graphs that have an
identical phase space structure. The relative factor of 2 between the two
types of factors is thus preserved.

%The simplest way to see that these diagrams do not
%contribute to the scalar potential is to note that the derivative interactions
%do not contribute to the 1-loop CW potential.

The diagrams of type (b) can be shown to sum to zero. This can be seen by
noting that for any given diagram where the gauge boson connects to an $H_L$
$(H_R)$ on the external leg, one can draw a different diagram in which the
gauge boson connects to $H_L^\dagger$ $(H_R^\dagger)$, yielding a relative
minus sign. The sum of all such diagrams cancels in pairs.

The contribution from diagrams of type~(c) are also trivially $U(4)$ symmetric
since the gauge boson loop is merely a correction to the propagator which
respects $U(4)$ due to the conserved $Z_2$ symmetry. $U(4)$ is thus not
violated by such diagrams.

The most non-trivial cancellation occurs in diagrams of type~(d). For brevity
we will simply quote the result here and present a more instructive proof in
the next appendix. The $|H_L|^4$ receives a contribution proportional to
$15\lambda^2 g^2$ (here the relative sign between some diagrams plays an
important role). The contribution to $|H_R|^2|H_L|^2$ is proportional to $30
\lambda^2 g^2$. The constants of proportionality in both cases are a common
phase space integral. The overall quartic from this class of diagrams is
thus $U(4)$ symmetric as well.

As mentioned above, the $U(1)_{B-L}$ contribution is $U(4)$ symmetric
because these two groups commute. However
in~\cite{twin} the $U(1)$ gauge structure is different. There two sets of
hypercharge, $A$ and $B$ were gauged. One can use the same arguments for
diagrams of type~(a)-(c) and explicitly calculate those of type (d) to find
that diagrams with hypercharge gauge boson exchange do not contribute an
$U(4)$ violating quartic either.

\section{Higher Order Corrections}\label{appendixB}

We now demonstrate that in the non-linear model there are no corrections of order $g^2
f^2$ to the mass of the pseudo-Goldstones at any order in perturbation theory. In
particular the cancellation of the two-loop diagrams in the previous section is not
accidental, but instead follows from a symmetry argument.

Consider first the linear model. We will show that no U(4) violating potential
terms are generated for $H$ at order $g^2$. We start first with U(1$)_{B - L}$.
At order $g'$ we have the following interaction between $H$ and $B_{\mu}$.
\begin{equation}
\frac{i}{2} g' B_{\mu} \; \left[ \partial^{\mu} H_L^{\dagger} \; H_L
+ \partial^{\mu} H_R^{\dagger} \; H_R
- H_L^{\dagger} \; \partial^{\mu} H_L -  H_R^{\dagger} \;
\partial^{\mu} H_R \right]
\end{equation}
We see that the couplings of $B_{\mu}$ at order $g'$ are invariant under the U(4)
symmetry under which $H = (H_L, H_R)$ transforms as a fundamental. Then the only terms
in the potential which can be generated from this interaction have the form of
$H^{\dagger} H$ raised to some power, which is U(4) invariant.

At order $g'^2$ we have the interaction
\begin{equation}
\frac{g'^2}{4} B'_{\mu} B'^{\mu} \left[ H_L^{\dagger} H_L + H_R^{\dagger} H_R
\right]
\end{equation}
which is also manifestly U(4) invariant and will not generate U(4)  violating
terms in the potential.

We now turn to the interactions of $H$ with the gauge bosons of SU(2$)_L$ and
SU(2$)_R$. Decompose $W_{{\mu}, L} = 1/2 \; W_{{\mu}, L}^a \tau^a$, $W_{{\mu}, R} =
1/2 \; W_{{\mu},R}^a \tau^a$ where $a$ runs from one to three and the $\tau^a$ are the
Pauli matrices. Since the gauge boson propagator is diagonal in the index $a$, to
order $g^2$ we are free to consider $a = 1$, $a = 2$ and $a = 3$ separately. For now
we therefore focus only on the interactions of $H$ with $W_{{\mu}, L}^3$ and
$W_{{\mu}, R}^3$. At order $g$ these take the form
\begin{eqnarray}
\label{orderg}
\frac{i}{2} \; g \; W_{\mu, L}^3 \; \left[ \partial^{\mu} H_L^{\dagger} \; \tau^3 \;
H_L - H_L^{\dagger} \; \tau^3 \; \partial^{\mu} H_L \right] \nonumber \\ + \frac{i}{2}
\; g \; W_{\mu, R}^3 \; \left[ \partial^{\mu} H_R^{\dagger} \; \tau^3 \;  H_R -
H_R^{\dagger} \; \tau^3 \; \partial^{\mu} H_R \right]
\end{eqnarray}
We can rewrite this interaction in terms of a new set of variables.  Expanding out
$H_L = (H_{L1}, H_{L2})$, $H_R = (H_{R1}, H_{R2})$ we can define
\begin{eqnarray}
H_{3, +} &=& \left( H_{L1},H_{R1},H^*_{L2},H^*_{R2} \right )\nonumber\\
H_{3, -} &=& \left( H_{L1},H_{R2},H^*_{L2},H^*_{R1} \right).
\end{eqnarray}
Further, define
\begin{eqnarray}
W_{\mu, +} &=& \frac{1}{{2}} \left[ W_{\mu, L} + W_{\mu, R} \right] \nonumber
\\
W_{\mu, -} &=& \frac{1}{{2}} \left[ W_{\mu, L} - W_{\mu, R} \right]
\end{eqnarray}
Note that the gauge boson propagators are diagonal in the
$W_{\mu, +}, W_{\mu, -}$ basis.
In terms of these new variables eq.~(\ref{orderg}) becomes
\begin{eqnarray}
\label{ordergagain}
\; \frac{i}{2} \; g \; W_{\mu, +}^3 \; \left[ \partial^{\mu} H_{3,+}^{\dagger} \;
H_{3,+} -  H_{3,+}^{\dagger} \; \partial^{\mu} H_{3,+} \right] \nonumber \\
+ \; \frac{i}{2} \; g \; W_{\mu, -}^3 \; \left[ \partial^{\mu} H_{3,-}^{\dagger} \;
H_{3,-} -  H_{3,-}^{\dagger} \; \partial^{\mu} H_{3,-} \right]
\end{eqnarray}
We see that the couplings of $W_{\mu, +}^3$ at order $g$ are invariant under
a U(4) symmetry under which $H_{3,+}$ transforms as a fundamental. Therefore
the only potential terms which can be generated from this interaction at
order $g^2$ have the form of $H_{3,+}^{\dagger} H_{3,+} = H^{\dagger}H $
raised to some power.  Similarly the couplings of $W_{\mu, -}^3$ at order $g$
are invariant under a different U(4) symmetry under which $H_{3,-}$
transforms as a fundamental. Again the only potential terms this allows at
order $g^2$ have the form of $H_{3,-}^{\dagger} H_{3,-} = H^{\dagger} H$
raised to some power. Although the argument we have just given applies only
to $W_{\mu, L}^3$ and $W_{\mu, R}^3$ it generalizes in a straightforward way
to the other components of the SU($2)_L$ and SU($2)_R$ gauge bosons, since
the different contributions can be related to each other through SU(2)
rotations.

We now consider the interactions of $H$ with $W_{{\mu}, L}$ and $W_{{\mu},
R}$ at order $g^2$. These take the form
\begin{equation}
g^2 H_L^{\dagger} W_{\mu, L} W^{\mu, L} H_L + g^2 H_R^{\dagger} W_{\mu, R} W^{\mu, R}
H_R
\end{equation}
We rewrite the relevant terms in this interaction in terms of $W_{\mu, +}$ and
$W_{\mu, -}$ as
\begin{eqnarray}
\label{ordergsquared}
g^2 H_L^{\dagger} \left( W_{\mu, +} W^{\mu, +} + W_{\mu, -} W^{\mu, -} \right) H_L
\nonumber \\
+ g^2 H_R^{\dagger} \left( W_{\mu, +} W^{\mu, +} + W_{\mu, -} W^{\mu, -} \right) H_R
\end{eqnarray}
where we have dropped `mixed terms' such as $H_L^{\dagger} W_{\mu, +} W^{\mu,
-} H_L$ which cannot contribute to the quartic at order $g^2$. From eq.
(\ref{ordergsquared}) we see that the remaining terms are invariant under a
U(4) symmetry under which $H = (H_L, H_R)$ transforms as a fundamental and
therefore only give rise to a U(4) invariant potential terms at order $g^2$.
This completes the proof that in the linear model U(4) violating potential
terms are not generated at order $g^2$ to any order in perturbation theory.

We now consider the effect of adding arbitrary non-renormalizable
interactions to the linear model. The additional terms are assumed to be
invariant under O(8), with the SU(2$)_L$, SU(2$)_R$ and U(1$)_{B - L}$
subgroups gauged. In general these new terms can be constructed by making
Lorentz invariant contractions or products from gauge invariants of the form
\begin{equation}
\left[D_{\alpha} D_{\beta} . . . H \right]^{\dagger}
\left[D_{\lambda} D_{\sigma} . . . H \right] \; + \;{\rm h.c. }
\end{equation}
where the number `n' of gauge covariant derivatives $D_{\alpha}^{\dagger}$
acting on $H^{\dagger}$ and the number `m' of gauge covariant derivatives
$D_{\lambda}$ acting on $H$ are both arbitrary. Hermitian conjugation is
necessary for O(8) invariance.
Let us first consider the
interactions between $H$ and the SU(2) gauge bosons at order $g$. We take as
a representative term
\begin{eqnarray}
[\partial^n H_L^{\dagger} \partial^{m-1} \left(ig W_L H_L \right)
&+& \partial^n H_R^{\dagger} \partial^{m-1} \left( ig W_R  H_R \right)]
\nonumber \\
&+& \left[ {\rm h.c. } \right]
\end{eqnarray}
where for simplicity we have suppressed all Lorentz indices. Once again, we
go to the $W_{\mu, +}, W^{\mu, -}$ basis and restrict our consideration to
the third component of the $W$'s. Then the interaction above can be rewritten
as
\begin{eqnarray}
\frac{ig}{2}[\partial^n H_{3,+}^{\dagger} \partial^{m-1} \left(
W_{\mu,+}^3
H_{3,+} \right) &+&
\partial^n H_{3,-}^{\dagger} \partial^{m-1} \left(
W_{\mu,-}^3 H_{3,-} \right) ] \nonumber
\\
&+& \left[ {\rm h.c. } \right]
\end{eqnarray}
We see that to order $g$ all the interactions of $W_{\mu,+}^3$ with $H_{3,+}$
are invariant under a U(4) symmetry as in the simple linear model. Similarly
all the interactions of $W_{\mu,-}^3$ with $H_{3,-}$ are also invariant under
a U(4) symmetry to order $g$. Then to order $g^2$ the only potential terms
that can be generated from these couplings have the form of the U(4)
invariant $H^{\dagger}H$ raised to some power.

We now consider the interactions between $H$ and the SU(2) gauge bosons at
order $g^2$. We take as a representative term
\begin{eqnarray}
[ \partial^n H_L^{\dagger} \partial^{m-2} \left(g W_L \right)^2 H_L
&+& \partial^n H_R^{\dagger} \partial^{m-2} \left( g W_R \right)^2 H_R]
 \nonumber \\
&+& \left[{\rm h.c. } \right]
\end{eqnarray}
This can be rewritten
\begin{eqnarray}
\left[ \partial^n H_L^{\dagger} \partial^{m-2} \left(g^2 W_+ W_+ + g^2
W_-W_-\right) H_L \right]
&+&
 \\
\left[ \partial^n H_R^{\dagger} \partial^{m-2} \left( g^2 W_+ W_+ + g^2 W_-W_-
\right) H_R
\right] &+& \left[ {\rm h.c. } \nonumber
\right]
\end{eqnarray}
where we have dropped the terms that involve products of $W_+$ and $W_-$
because they do not contribute at order $g^2$. The remaining terms are
invariant under a U(4) symmetry under which $H = (H_L, H_R)$ transforms as a
fundamental and therefore only give rise only to U(4) invariant terms at
order $g^2$, just as in the simple linear model.

We have therefore shown that the SU(2) gauge interactions do not generate a
U(4) violating terms in the potential to order $g^2$ at any order in
perturbation theory.  
%%%%
One can summarize this proof as follows. Every gauge generator $\tau^{a}_{\pm}$
breaks the global $O(8)$ symmetry in the Higgs interactions down to a global
SU(4) subgroup, which is different for every generator. Every such $SU(4)$ is
enough to forbid the $|H_L|^4 + |H_R|^4$.  Therefore, in order to generate
this quartic two different gauge generators must be involved. However, since
the different gauge bosons do not mix via the kinetic terms, any contribution
that involves two generators will be proportional to~$g^4$.
%%%
It is straightforward to include $U(1)_{B - L}$ in this argument
using the same methods. It follows that adding arbitrary non-renormalizable
interactions to the linear sigma model does not result in $U(4)$ violating
terms in the potential at order $g^2$. Since the general non-linear model
with the symmetry properties we desire may be obtained by integrating out the
radial mode from this model, it follows that in the non-linear model the
pseudo-Goldstones do not acquire a mass at order $g^2$ to any order in
perturbation theory. Although the pseudo-Goldstones do acquire a mass at
order $g^4$ from states at the cutoff this is always further loop suppressed,
and is expected to be smaller than the logarithmically enhanced contribution
calculated in the body of the paper.  The very same arguments can be applied
to the mirror twin Higgs model of ref~{\cite{twin}} to establish the absence
of corrections to the pseudo-Goldstone potential at order $g^2$.

\end{document}